\documentclass[aps,twocolumn,english,pra,showpacs]{revtex4}
\usepackage{amsmath}
\usepackage{amssymb}
\usepackage{graphicx}

\providecommand{\tabularnewline}{\\}

\usepackage[pdfview=FitH,colorlinks=true,linkcolor=blue,citecolor=blue,urlcolor=blue]{hyperref}
\allowdisplaybreaks

\usepackage{babel}
\begin{document}

\title{Instabilities of Bosonic Spin Currents in Optical Lattices}

\author{Hoi-Yin Hui, Ryan Barnett, Rajdeep Sensarma, S. Das Sarma}

\affiliation{Condensed Matter Theory Center and Joint Quantum Institute, Department
of Physics, University of Maryland, College Park, Maryland 20742,
USA}
\begin{abstract}
  We analyze the dynamical and energetic instabilities of spin
  currents in a system of two-component bosons in an optical lattice,
  with a particular focus on the Neel state. We consider both the weakly
  interacting superfluid and the strongly interacting Mott insulating
  limits as well as the regime near the superfluid-insulator
  transition and establish the criteria for the onset of these
  instabilities. We use Bogoliubov theory to treat the weakly
  interacting superfluid regime. Near the Mott transition, we
  calculate the stability phase diagram within a variational
  Gutzwiller wavefunction approach. In the deep Mott limit we discuss
  the emergence of the Heisenberg model and calculate the stability
  diagram within this model. Though the Bogoliubov theory and the Heisenberg
  model (appropriate for deep superfluid and deep Mott phase
  respectively) predict no dynamical instabilities, we find,
  interestingly, between these two limiting cases there is a regime of
  dynamical instability. This result is relevant for the ongoing
  experimental efforts to realize a stable Neel-ordered state in
  multi-component ultracold bosons.
\end{abstract}


\maketitle

\section{Introduction}

Ultracold atomics gases have recently emerged as a very important
platform to study non-equilibrium quantum dynamics of interacting
many-body systems. The tunability of Hamiltonian parameters together
with almost complete isolation from the environment and the long
time-scales in these systems have made it possible to study intrinsic
non-equilibrium dynamics of these systems without ultrafast probes.
Recently, there has been a growing number of experimental and
theoretical investigations of the dynamical properties of
Bose-Einstein Condensates (BECs) in optical lattices
\cite{Morsch2006Dynamics}.  Of particular interest  are experiments
which exhibit a \emph{dynamical instability}, which is a generic
phenomenon present in nonlinear systems under appropriate conditions.
Previously established examples of dynamical instabilities occur in
water waves \cite{Whitham1965general,Benjamin1967Instability}, light
in dielectric media
\cite{Ostrovskii1964,Ostrovskii1967,Hasegawa1980Tunable,Tai1986Observation},
 and plasmas
\cite{Hasegawa1972Theory,Pesme1988Modulational,McKinstrie1989modulational}. Recently, dynamical instabilities have been observed in ultracold fermi gases after a tuning of the interaction parameters~\cite{Joexpt,Pekkertheory}.
The realization of dynamical instabilities for current-carrying states
in BECs have received considerable attention both theoretically
\cite{Wu2001Landau,Konotop2002Modulational,Smerzi2002Dynamical,Machholm2003Band,Modugno2004Role,Altman2005Superfluid-Insulator,Polkovnikov2005Decay}
as well as experimentally
\cite{Raman1999Evidence,Cataliotti2003Superfluid,Fallani2004Observation,Anker2005Nonlinear,Fertig2005Strongly,DeSarlo2005Unstable,Mun2007Phase,Ferris2008Dynamical}.

Two qualitatively distinct types of instabilities can occur for
interacting systems of bosons: (i) the energetic instability and (ii)
the dynamical instability.  The energetic instability occurs if the system
is not at a local minimum of the mean-field energy.  If the system is
capable of dissipating energy, then it will decay from the initial metastable
state thereby exhibiting the instability.  A well-known
example for this case is the Landau instability (LI) for which a
superfluid carrying current in excess of the sound velocity becomes
unstable, leading to a breakdown of superfluidity.  In contrast,
a dynamical instability (DI) occurs when the system has collective
modes with complex frequencies. Such modes will
result in an exponential growth of small perturbations which manifests
as a rapid depletion and fragmentation of the condensate
\cite{Fallani2004Observation,Ferris2008Dynamical}.  For systems that
do not have a dissipative mechanism, the energetic instability alone
will not occur.  On the other hand, the dynamical instability occurs
even without dissipation, and will be observable unless the growth
time of the most unstable mode is longer than experimental time
scales.  It can also be seen that an energetic instability is a
necessary condition for a dynamical instability.

Bosons in an optical lattice undergo a quantum phase transition from a
superfluid phase in the weakly interacting limit to an incompressible
Mott insulator phase as the interaction parameter increases beyond a
critical value. Scalar bosonic condensates have a U(1) symmetry
associated with the superfluid phase, resulting in a conserved mass
current.  In the presence of a lattice, when the externally imposed
current exceeds a critical value, the system manifests a dynamical
instability~\cite{Wu2001Landau,Konotop2002Modulational,Smerzi2002Dynamical}.
The critical current required for the dynamical instability decreases
with increasing interaction strength and vanishes at the critical
interaction required for the superfluid-insulator transition.
Additional types of dynamical instabilities can occur in
multicomponent condensates due the their more complex order
parameters. In this paper we will focus on two-component bosons in an
optical lattice with spin-independent interactions. In addition to the
superfluid-insulator transition, this system also shows a spontaneous
ferromagnetic spin ordering in the equilibrium ground state.  This
system has an SU(2) symmetry due to invariance of the energy under
spin rotation which results in a conserved spin current. In the
presence of externally imposed spin currents (spin twists), the system
exhibits dynamical instabilities when the spin current exceeds a
critical value. We will mainly focus on these spin-current driven
instabilities, which occur in addition to and even in the absence of
any mass current driven instabilities.

Previous work addressing spin current instabilities in bosonic systems
have focused on the continuum, weakly interacting superfluids where
the Gross-Pitaevskii Equation (GPE) is applicable.  In such a context,
the counterflow instability
\cite{Law2001Critical,Kuklov2003Counterflow,Takeuchi2010Binary,Hoefer2010Counterflow,Hamner2011Generation}
as well as the instability of a spin-one condensate from an initial
helical state \cite{Vengalattore08,Cherng2008Dynamical} have been
investigated.  Here we analyze the instabilities of the system in the
presence of an optical lattice for a wide range of interaction
parameters going from the weakly interacting limit (the deep
superfluid phase) through the intermediate regime near the
superfluid-insulator transition to the strongly interacting (atomic)
limit, deep into the Mott phase. The weakly interacting regime is
treated within the standard Bogoliubov theory, while the strongly
interacting regime is treated within a spin-wave approximation of the
ferromagnetic Heisenberg model, where the spin-spin interaction comes
from super-exchange mechanism. The intermediate interaction regime is
treated within a variational Gutzwiller wavefunction ansatz. We extend
the Gutzwiller ansatz to both the deep superfluid and the deep Mott
limit and compare the results with those from the more established
formalisms mentioned above.

To analyze stability of the bosonic states, we construct either mass
or spin current carrying mean-field states. The spectrum of quantum
fluctuations about these stationary states is then calculated within a
Gaussian approximation. Negative eigenvalues of the fluctuation
Hamiltonian indicate an energetic instability while a complex collective
mode spectrum indicates a dynamical instability. For a dynamical
instability, the positive imaginary part of the complex spectrum gives
the growth rate of the unstable fluctuation modes. Our main results
are: (i) We show that the mass current induced instabilities give rise
to the same instability phase diagram in the critical current
interaction plane for both spinless and two-component bosons. (ii) The
two-component bosons exhibit a spin-current induced dynamical
instability in a large region of the critical current interaction
strength plane in the superfluid phase. We also show the collective
modes which are unstable and compute their growth rates. (iii) We
focus on the Neel-ordered state, which can be interpreted as a
spin-current-carrying state with particular commensurate wavevector.
Although the Neel configuration is not the ground state of the system,
there are proposals~\cite{Sorensen2010Adiabatic} to experimentally
explore the physics about this high energy state provided its lifetime
is sufficiently long.  We show that while this state is stable in the
deep superfluid and insulating limits, in the intermediate regime,
interestingly, the system is dynamically unstable. We thus provide a
comprehensive picture of the spin-current induced dynamical
instabilities in two-component bosons on optical lattices for a wide
range of interactions and spin currents.

The paper is organized as follows. In Sec.~\ref{sec:chargecurrent}, we
review the established DI of the mass current of spinless bosons.  We
use the Bogoliubov theory to analyze the superfluid limit and the
Gutzwiller ansatz to analyze the strongly-interacting regime close to
Mott boundary.  This prepares us to investigate the instabilities
related to the spin current of a two-component Bosonic condensate in
Sec.~\ref{sec:spincurrent} in the regime of weak as well as
intermediate interactions.  We shall present the stability phase
diagram and discuss how our results are connected to the deep Mott
limit.  In Sec.~\ref{sec:Discussion} we discuss the stability of the
Neel state limit for different regimes.  Finally in
Sec.~\ref{sec:Conclusion} we summarize our results.

\section{Instabilities of moving scalar
  condensates \label{sec:chargecurrent}}

For completeness and to set the notation and general approach, we
first briefly consider the mass current in a single-component BEC and
the concomitant Landau and dynamical instabilities. The weakly
interacting superfluid case was originally considered in
Refs.~\cite{Smerzi2002Dynamical,Wu2001Landau}, while the regime near
the Mott transition was addressed in
Ref.~\cite{Altman2005Superfluid-Insulator, Polkovnikov2005Decay}.  A system of bosons on a
lattice and in the lowest band is described by the Bose-Hubbard model
\begin{equation}
H=-t\sum_{\left\langle ij\right\rangle }\left(b_{i}^{\dagger}b_{j}+\mathrm{hc}\right)+\frac{U}{2}\sum_{i}\left(n_{i}-\bar{n}\right)^{2}-\mu\sum_{i}\left(n_{i}-\bar{n}\right)\label{eq:1BH}
\end{equation}
where $b^\dagger_i$ is the boson creation operator on the lattice site
$i$, $t$ is the hopping matrix element, $U$ is the on-site repulsion,
$\mu$ is the chemical potential, $n_{i}=b^{\dagger}_i b_i$, and
$\bar{n}$ is the average number of bosons per site.  We consider this
model in one, two, and three dimensions for cubic lattices.  When
$t\gg U/\bar{n}$ the system has a superfluid ground state, and
Bogoliubov theory describes its elementary excitations.  When $t \sim
U/\bar{n}$, there is a quantum phase transition at $U=U_c$ to an incompressible
Mott state. The Bogoliubov theory fails in the vicinity of this
transition, however, a variational Gutzwiller ansatz can be used to
treat the system in this regime.

\subsection{Weakly Interacting Superfluid}

Deep in the superfluid phase the current-carrying states can be represented by a condensate wavefunction of the form
\begin{equation}
\bar{b}_{i}=\sqrt{n}e^{i\mathbf{p}\cdot\mathbf{x}_{i}}.
\end{equation}
which has a phase twist along $\hat{p}$ and carries a mass current
$\propto\sin \bf{p\cdot\bf{x_{ij}}}$ between neighboring sites.  This
wavefunction can be found within mean-field theory by solving the
Gross-Pitaevskii equation.  Expanding the energy of the system,
Eq.~(\ref{eq:1BH}), about this state to quadratic order, with
$b_i=\bar{b}_i+\phi_i$, one obtains the fluctuation Hamiltonian
$\delta
H=\sum_{\mathbf{k}}\Phi_{\mathbf{k}}^{\dagger}M_{\mathbf{k}}(\mathbf{p})\Phi_{\mathbf{k}}$
where
$\Phi_{\mathbf{k}}^{\dagger}=\left(\phi_{\mathbf{k}}^{\dagger},\phi_{-\mathbf{k}}\right)$
and
\begin{equation}
M_{\mathbf{k}}=\left(\begin{array}{cc}
\epsilon_{\mathbf{k}+\mathbf{p}}-\epsilon_{\mathbf{p}}+Un & Un\\
Un & \epsilon_{\mathbf{k}-\mathbf{p}}-\epsilon_{\mathbf{p}}+Un
\end{array}\right)
\end{equation}
with $\epsilon_{\mathbf{q}}=-zt\gamma_{\mathbf{q}}$, where $z$ is
the coordination number and $\gamma_{\mathbf{q}}=z^{-1}\sum_{\mathbf{\delta}}e^{i\mathbf{q}.\mathbf{\delta}}$.
The energies of the normal modes of the system are given by the eigenvalues of the
matrix $\sigma_z M_{\mathbf{k}}$ \cite{Wu2001Landau} where $\sigma_z$
is a Pauli matrix.  On the other
hand, if the system is at a local minimum in energy, then the matrix
$M_{\mathbf{k}}$ itself will be positive definite.
We thus summarize the following criteria
for the instabilities:
\begin{itemize}
\item LI: at least one eigenvalue of $M_{\mathbf{k}}$ is negative
\item DI: at  least one eigenvalue of $\sigma_{z}M_{\mathbf{k}}$ is complex
\end{itemize}
For mass-current-carrying states, it is well known that the continuum
theory only sustains Landau instabilities, which occur when the
current in the system exceeds the speed of sound. There are no
dynamical instabilities in the continuum theory. However, on a lattice
the system exhibits both Landau and dynamical instabilities with the
criterion for critical current summarized in Table~\ref{tab:Bogo}. The
dynamical instability is crucially related to the softening of
collective modes at finite wavevectors, which does not occur in the
continuum.

\begin{table}
\hfill{}%
\begin{tabular}{|c||c|c||c|c|}
\hline 
\multicolumn{1}{|c||}{} & \multicolumn{2}{c||}{Phase twist} & \multicolumn{2}{c|}{Spin Twist}\tabularnewline
\hline 
 & Continuum & Lattice & \multicolumn{1}{c|}{Continuum} & \multicolumn{1}{c|}{Lattice}\tabularnewline
\hline 
\hline 
LI & $\tilde{p}>\sqrt{mUn}$ & $\frac{\sin^{2}\tilde{p}}{\cos \tilde{p}}>\frac{Un}{zt}$ & \multicolumn{1}{c|}{$\tilde{p}\neq0$} & \multicolumn{1}{c|}{$\tilde{p}\neq0\,\mathrm{mod}\,2\pi$}\tabularnewline
\hline 
DI & Never & $\cos \tilde{p}<0$ & \multicolumn{1}{c|}{$\tilde{p}\neq0$} & \multicolumn{1}{c|}{$\tilde{p}\neq0,\pm\frac{\pi}{2}\,\mathrm{mod}\,2\pi$}\tabularnewline
\hline 
\end{tabular}\hfill{}

\caption{ The conditions for Landau and dynamical instabilities of
mass-current carrying states (for spinless bosons) and spin-current
carrying states (for 2-component spin-full bosons) in the weakly
interacting limit on a square lattice (calculated within Bogoliubov
theory).  The results are valid for one, two, and three dimensions.
The case of twisting along the diagonal of the square lattice:
$\mathbf{p}=\tilde{p}\sum_{i}\hat{x}_{i}$ is taken.  For comparison,
the conditions for the instabilities in the continuum are also given.}
\label{tab:Bogo}
\end{table}

\subsection{Gutzwiller Ansatz}

\begin{figure}
\begin{centering}
\includegraphics[width=5cm]{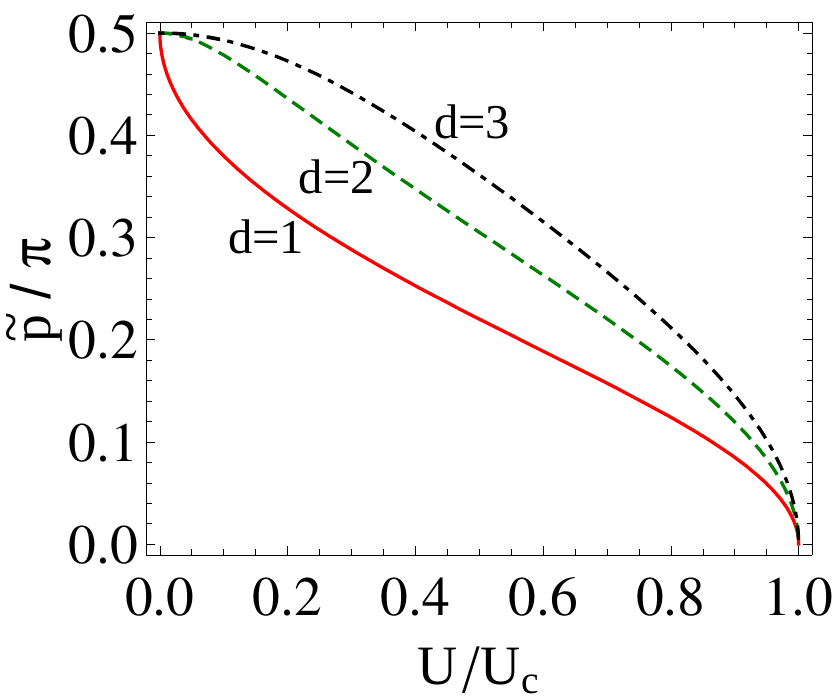}
\par\end{centering}

\caption{Stability phase diagram of phase twists in a lattice in different
dimensions, where twisting is imposed along: $\mathbf{p}=\tilde{p} \hat{x}$. The same diagram is obtained for
either spinless bosons Eq.~(\ref{eq:1BH}) or the two-component bosons
Eq.~(\ref{eq:2BH}). This diagram agrees well with the more extensive
treatment presented in
\cite{Altman2005Superfluid-Insulator,Polkovnikov2005Decay}.
\label{fig:spinlessPhase}}
\end{figure}

To investigate the DI for the Bose-Hubbard model for stronger
interactions, we shall approach the problem within a truncated Hilbert
space. We consider the variational Gutzwiller wavefunction for the
ground state, $\left|\Omega\right\rangle
=\prod_{i}\left|\Omega_{i}\right\rangle $, with
$\left|\Omega_{i}\right\rangle
=\cos\frac{\theta_{i}}{2}\left|\bar{n}\right\rangle _{i}+
e^{i\eta_{i}}\sin\frac{\theta_i}{2}\left[\cos\frac{\chi_i}{2}e^{-i\varphi_{i}}\left|\bar{n}-1\right\rangle_{i}+\sin\frac{\chi_{i}}{2}e^{i\varphi_{i}}\left|\bar{n}+1\right\rangle_{i}\right]$,
where $\left|n\right\rangle_i$ are the Fock states on the site
$i$. This variational state was used in
Ref.~\cite{Altman2002Oscillating} to study the Bose-Hubbard model near
the Mott transition in the absence of a current. Our calculations
follow along similar lines with an important distinction: the phase
$\varphi_i$ is position-dependent, i.e.
$\varphi_{i}=\mathbf{p}\cdot\mathbf{x}_{i}$ such that $\left\langle
  \Omega\right|b_{i}\left|\Omega\right\rangle \propto\sin\theta
e^{i\mathbf{p}\cdot\mathbf{x}_{i}}$, which ensures that the state
carries a mass current flowing along $\hat{p}$. Other parameters are
then varied to minimize the energy of this mean-field state, giving
$\chi=\frac{\pi}{2}$ and $\eta=0$.

The Hamiltonian is expanded about this stationary state in the following
way: we introduce the bosonic pseudospin operators $t_{\alpha i}^{\dagger}\left|\mathrm{vac}\right\rangle \equiv\left|\bar{n}+\alpha\right\rangle _{i}$,
$\alpha=\{-1,0,1\}$ with the constraint $\sum_{\alpha}t_{\alpha i}^{\dagger}t_{\alpha i}=1$,
so that the Boson operators can be written as $b_{i}^{\dagger}\rightarrow\sqrt{\bar{n}}t_{0i}^{\dagger}t_{-1i}+\sqrt{\bar{n}+1}t_{1i}^{\dagger}t_{0i}$.
A unitary transformation is then performed with
\begin{equation}
\left(\begin{array}{c}
d_{0i}^{\dagger}\\
d_{1i}^{\dagger}\\
d_{2i}^{\dagger}
\end{array}\right)=\left(\begin{array}{ccc}
\frac{e^{-i\mathbf{p}\cdot\mathbf{x}_{i}}}{\sqrt{2}}\sin\frac{\theta}{2} & \cos\frac{\theta}{2} & \frac{e^{i\mathbf{p}\cdot\mathbf{x}_{i}}}{\sqrt{2}}\sin\frac{\theta}{2}\\
-\frac{e^{-i\mathbf{p}\cdot\mathbf{x}_{i}}}{\sqrt{2}}\cos\frac{\theta}{2} & \sin\frac{\theta}{2} & -\frac{e^{i\mathbf{p}\cdot\mathbf{x}_{i}}}{\sqrt{2}}\cos\frac{\theta}{2}\\
\frac{e^{-i\mathbf{p}\cdot\mathbf{x}_{i}}}{\sqrt{2}} & 0 & -\frac{e^{i\mathbf{p}\cdot\mathbf{x}_{i}}}{\sqrt{2}}
\end{array}\right)\left(\begin{array}{c}
t_{-1i}^{\dagger}\\
t_{0i}^{\dagger}\\
t_{1i}^{\dagger}
\end{array}\right)\label{eq:spinlessUnitary}
\end{equation}
and the Hamiltonian is written in terms of the $d$ operators. Since
$d_{0}^{\dagger}\left|\mathrm{vac}\right\rangle $ represents the
minimum energy state, it is macroscopically occupied, while $d_{n>0}^{\dagger}$
are fluctuations about this state. Therefore, we eliminate
$d_{0}^{\dagger}$ using $d_{0}^{\dagger}\approx d_{0}\approx1-\frac{1}{2}d_{1i}^{\dagger}d_{1i}-\frac{1}{2}d_{2i}^{\dagger}d_{2i}$,
which resembles the Holstein-Primakoff transformation
\cite{Holstein1940Field} used in spin models.

To the quadratic order in the operators $d_{n>0}^{\dagger}$, the
Hamiltonian has the form $H=\Psi^{\dagger}_{{\bf k}}M_{{\bf
    k}}\Psi_{{\bf k}}$ where $\Psi^{\dagger}_{{\bf
    k}}\equiv\left(d_{1k}^{\dagger},d_{1,-k},d_{2k}^{\dagger},d_{2,-k}\right)$
and the form of $M_{{\bf k}}$ is given in
Appendix~\ref{sec:App-spinless}. For a given $U/t$ and $\mathbf{p}$, we
compute the energies $\omega_{1,2k}$ for $k\in\left[-\pi,\pi\right]$
by a Bogoliubov transformation.  As noted before, the presence of
complex eigenfrequencies indicate a dynamical instability.

For direct comparison with previous work, we consider $\mathbf{p}$
along an axis of a $d$-dimensional cubic lattice $\left(\mathbf{p}=p\hat{x}_1\right)$, giving $\gamma_\mathbf{p}=\frac{\cos p+(d-1)}{d}$.
The resulting phase diagram is shown in Fig.~\ref{fig:spinlessPhase},
which shows good agreement with the results in
\cite{Altman2005Superfluid-Insulator,Polkovnikov2005Decay} where a numerical analysis is
performed, taking a larger Hilbert space.

The Bogoliubov analysis is justified only if the fluctuation
occupation $\left\langle
d_{1i}^{\dagger}d_{1i}+d_{2i}^{\dagger}d_{2i}\right\rangle $ is
small compared to unity. This is checked in the stable regimes after the Bogoliubov
transformation is done. We find that for the 2D system, the fluctuation is
less than $0.2$ for all $U>0.1U_{c}$ and reach up to $0.5$ as
$U\rightarrow0$ and $p\rightarrow\pi/2$. This means the quantitative
result should be trustworthy for $U>0.1U_{c}$. However in 1D we always
find divergent occupation of the fluctuations as expected because of
the significance of quantum fluctuations. The qualitatively good
agreement for 1D results with experiment might be understood as due to
the logarithmic nature of the divergence, which is not severe in
finite-sized systems.

\section{Spin Current Instabilities in Two-Component Condensates
\label{sec:spincurrent}}
Having set up the formalism to study Landau and dynamical
instabilities in spinless bosonic systems, we will now adapt this formalism to
study instabilities of spin-current carrying states in condensates of
two-component bosons.  The starting
point for our analysis is the two-component rotationally invariant
Bose-Hubbard model
\begin{eqnarray}
H & = & -t\sum_{\left\langle ij\right\rangle \sigma}\left(b_{i\sigma}^{\dagger}b_{j\sigma}+\mathrm{hc}\right)+\frac{U}{2}\sum_{i}\left(n_{i}-\bar{n}\right)^{2}\nonumber \\
 &  & -\mu\sum_{i}\left(n_{i}-\bar{n}\right)\label{eq:2BH}.
\end{eqnarray}
where $b^\dagger_{i\sigma}$ creates a boson of spin $\sigma$ on site
$i$, $n_i = \sum_{\sigma}b_{i\sigma}^\dagger b_{i \sigma}$, and
$\bar{n}$ is the average particle number per site.  Such a
system could be realized using, for instance, two hyperfine states of
alkali atoms.  Due to the smallness of spin-exchange interaction for
typical alkali atoms, such systems possess an approximate SU(2)
symmetry, which is reflected in the spin-independent form of the
interactions which we consider here.   For simplicity we will
concentrate
on the case when $\bar{n}=1$ except for the Bogoliubov analysis.

The weakly interacting superfluid phase (first considered in
Ref.~\cite{Law2001Critical}) of the spinfull bosons is described by the
Bogoliubov theory around a mean-field state with a two-component
condensate wavefunction.  The intermediate interaction regime near the
Mott transition is analyzed, as before, with a variational Gutzwiller
ansatz, albeit with an extended local Hilbert space. However, unlike
the spinless bosons, the ferromagnetic spin-spin interaction in the
deep Mott phase, arising out of a super-exchange mechanism, is not
captured by the simple Gutzwiller ansatz. To treat this limit, we work
with a ferromagnetic Heisenberg model with a spin-spin interaction
$J=4t^2/U$ and analyze the spin-current induced instabilities within a
spin-wave formalism.

\subsection{Weakly interacting Superfluid\label{sec:spinfulGP}}

The weakly interacting superfluid regime 
admits coherent mean-field spin current-carrying solutions of the form
\begin{equation}
\mathbf{\bar{b}}_{i}=e^{i\sigma^{x}\mathbf{p}\cdot\mathbf{x}_{i}/2}\left(\begin{array}{c}
\sqrt{n}\\
0
\end{array}\right).
\end{equation}
Such states have a spin twist of ${\bf p}$, and carry spin current
$\propto\sin \mathbf{p}\cdot\mathbf{x}_{ij}$ between neighbors.
Expanding Eq.~(\ref{eq:2BH}) about this stationary state to second
order in quantum fluctuations,
$b_{i\sigma}=\bar{b}_{i\sigma}+\phi_{i\sigma}$, gives the Hamiltonian
$\delta
H=\sum_{\mathbf{k}>0}\Phi_{\mathbf{k}}^{\dagger}M_{\mathbf{k}}\Phi_{\mathbf{k}}$
where $\Phi_{\mathbf{k}}^{\dagger}=\left(
  \phi_{\uparrow\mathbf{k}}^{\dagger},\phi_{\uparrow -\mathbf{k}},\phi_{\downarrow\mathbf{k}}^{\dagger},\phi_{\downarrow-\mathbf{k}}\right)$
and
\begin{eqnarray}
M_{\mathbf{k}} & = & \left(\begin{array}{cccc}
\xi+2Un & Un & \epsilon_{-} & 0\\
Un & \xi+2Un & 0 & -\epsilon_{-}\\
\epsilon_{-} & 0 & \xi+Un & 0\\
0 & -\epsilon_{-} & 0 & \xi+Un
\end{array}\right)
\end{eqnarray}
where
\begin{eqnarray*}
\xi & = & \epsilon_{+}-\mu\\
\epsilon_{\pm} & = & \frac{\epsilon_{\mathbf{k}+\mathbf{p}/2}\pm\epsilon_{\mathbf{k}-\mathbf{p}/2}}{2}
\end{eqnarray*}
with $\epsilon_{\mathbf{q}}=-zt\gamma_{\mathbf{q}}=-2t\sum_{i}\cos q_{i}$.
For given $U$ and $\mathbf{p}$, negative eigenvalues of $M_{\mathbf{k}}$
for some $\mathbf{k}$ indicates LI while imaginary eigenvalues of
$\sigma M_{\mathbf{k}}$ indicates DI, where $\sigma=\mathrm{diag}\left(1,-1,1,-1\right)$.

From here on, we will restrict ourselves to the case of spin
currents along the diagonal of a square lattice:
$\mathbf{p}=\tilde{p}\left(\hat{x}+\hat{y}\right)$ (for example
$\tilde{p}=\pi$ represents the Neel state). The conditions for
instabilities are summarized in Table~\ref{tab:Bogo}. The LI is always
present for any non-zero pitch, while DI is always present except for
the $\tilde{p}=0$ ferromagnetic state and the $\tilde{p}=\pi$ Neel state.

To obtain a better understanding of the DI, we plot the wavevectors of
the unstable modes, obtained from the Bogoliubov theory, as a function
of the spin twist $\tilde{p}$ in the left column of
Fig.~\ref{fig:pkphase}. Here we consider wavevectors parallel to spin
current ($\mathbf{k}\parallel\mathbf{p}$) for several values of
$U$. Light gray areas indicate presence of LI but not DI, and dark
areas indicate the presence of both LI and DI. The ferromagnetic state
is always energetically and dynamically stable, as expected, while the
Neel state has a LI but not a DI.  With increasing $U$ the region
where the DI is present increases, i.e. more and more wavevectors
become unstable. However, the region where LI is present is almost
independent of $U$.

The dispersions of the lowest collective modes
($\mathbf{k}\parallel\mathbf{p}$) for three special states, the
ferromagnetic state ($\tilde{p}=0$), the Neel state ($\tilde{p}=\pi$)
and the spin spiral state with a wavelength of $4$ lattice spacings
($\tilde{p}=\pi/4$), are plotted in the left column of
Fig.~\ref{fig:Dispersions}. For the ferromagnetic state, there are
two low energy modes: a charge mode related to the U(1) symmetry
breaking, which disperses linearly and a spin mode related to the
SU(2) symmetry breaking, which disperses quadratically, both of which
are stable modes. As a spin current is imposed, the charge mode
remains stable while the spin mode develops a DI near $k=0$, indicated
by the thin red line.  As we reach the Neel state, both the charge and
the spin mode disperse linearly and are stable. Thus the DI disappears
for the Neel state, which is stable in the weakly interacting
limit. However, states with spin twists close to but not equal to
$\pi$, are unstable with the instability being seeded around the
wavevector $\mathbf{k}=\pi$.

\subsection{Gutzwiller Ansatz \label{sec:spinfulGutz}}

\begin{figure}
\begin{centering}
\includegraphics[width=7cm]{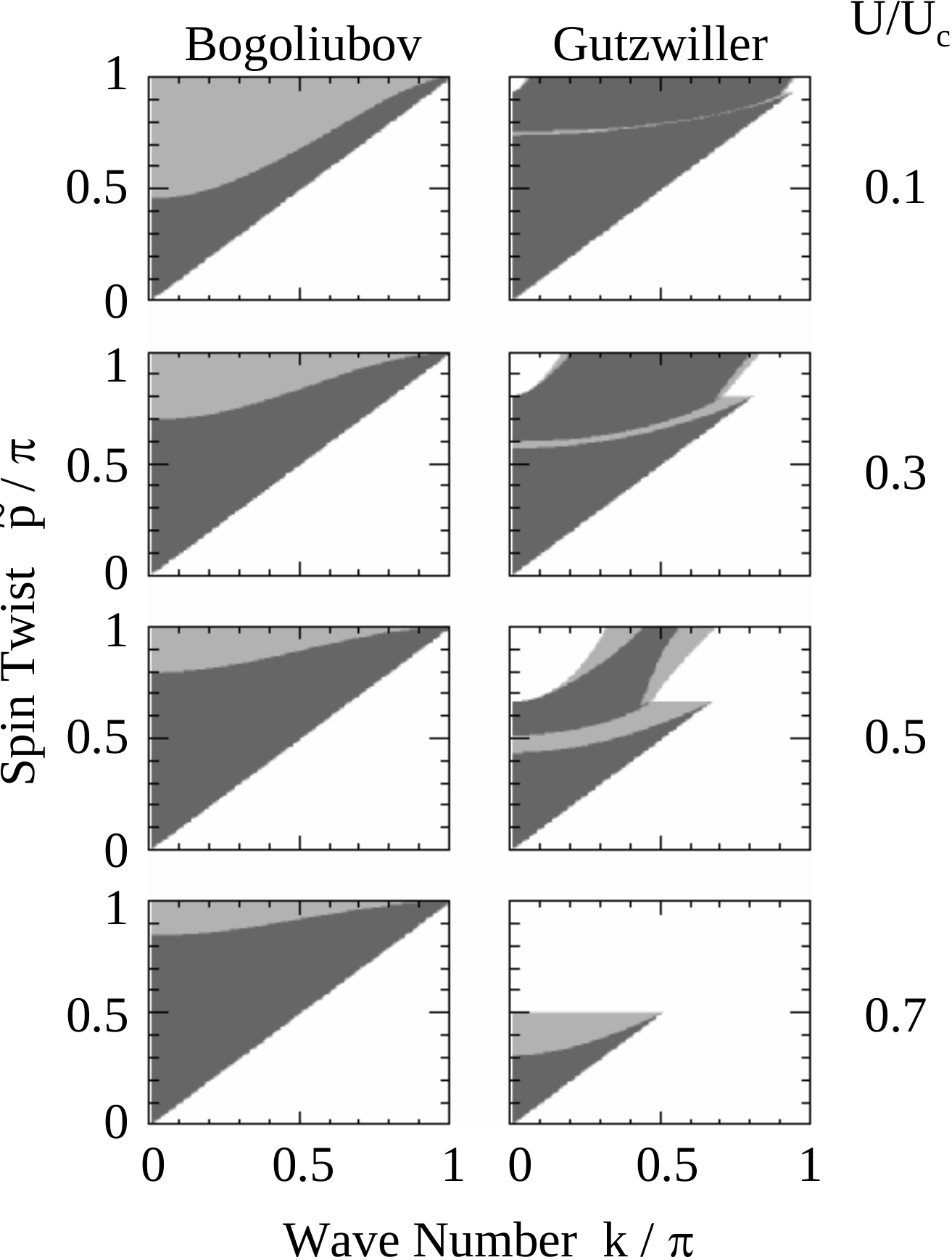}
\par\end{centering}

\caption{Diagrams showing regions of unstable modes for BEC spin-current-carrying
state in optical lattice, analyzed with Bogoliubov theory (left column) and Gutzwiller ansatz (right column). The value of $U/U_{c}$ used is (from top to
bottom) $0.1$, $0.3$, $0.5$, $0.7$. The spin twist plotted is
from ferromagnetic state (${\bf p}=(0,0)$) to Neel state (${\bf p}=(\pi,\pi)$). $k$
is the wave vector of the normal mode.  The shaded area indicates a LI
(negative excitation energy), while dark shaded area indicates a DI
(complex frequencies). The left column was previously obtained in
Ref.~\cite{Law2001Critical}. \label{fig:pkphase}}
\end{figure}

To investigate the regime near the insulator-superfluid transition we
adopt the Gutzwiller approach of Sec.~\ref{sec:chargecurrent}.  For
simplicity, we restrict ourselves to the case of unity filling.  Then
there are minimally six basis states per site that need to be included
in the local Hilbert space: $\left\{ \left|0\right\rangle
  ,\left|\uparrow\right\rangle ,\left|\downarrow\right\rangle
  ,\left|\uparrow\uparrow\right\rangle
  ,\left|\uparrow\downarrow\right\rangle
  ,\left|\downarrow\downarrow\right\rangle \right\} $, where the last
three states have double occupancy. The local Gutzwiller wavefunction
is then parametrized in terms of ten variables (site indices omitted)
\begin{eqnarray}
  \left|\Omega\right\rangle  & = & \sin\frac{\theta}{2}\cos\frac{\theta_{2}}{2}\left|0\right\rangle +e^{i(\varphi_{0}+\varphi_{1})}\cos\frac{\theta}{2}\cos\frac{\theta_{1}}{2}\left|\uparrow\right\rangle \nonumber \\
  &  & +e^{i(\varphi_{0}-\varphi_{1})}\cos\frac{\theta}{2}\sin\frac{\theta_{1}}{2}\left|\downarrow\right\rangle \nonumber \\
  &  & +e^{2i(\varphi_{2}+\varphi_{4})}\sin\frac{\theta}{2}\sin\frac{\theta_{2}}{2}\cos\frac{\theta_{3}}{2}\cos\frac{\theta_{4}}{2}\left|\uparrow\uparrow\right\rangle \nonumber \\
  &  & +e^{2i(\varphi_{2}+\varphi_{3})}\sin\frac{\theta}{2}\sin\frac{\theta_{2}}{2}\sin\frac{\theta_{3}}{2}\left|\uparrow\downarrow\right\rangle \nonumber \\
  &  & +e^{2i(\varphi_{2}-\varphi_{4})}\sin\frac{\theta}{2}\sin\frac{\theta_{2}}{2}\cos\frac{\theta_{3}}{2}\sin\frac{\theta_{4}}{2}\left|\downarrow\downarrow\right\rangle \label{eq:spinfulGutz}.
\end{eqnarray}

A phase twist $\left(\left\langle b_{\sigma}\right\rangle \propto
  e^{i\mathbf{p}\cdot\mathbf{x}}\right)$ can be imposed by setting
$\varphi_{0}=\varphi_{2}=\mathbf{p}\cdot\mathbf{x}$ and other
$\varphi_{i\neq0,2}$ to be uniform; while a spin twist
$\left(\left\langle S^{+}\right\rangle \propto
  e^{i2\mathbf{p}\cdot\mathbf{x}}\right)$ can be imposed by setting
$\varphi_{1}=\varphi_{4}=\mathbf{p}\cdot\mathbf{x}/2$ and other
$\varphi_{i\neq1,4}$ to be uniform. Note that, in our parametrization,
the spin-current-carrying states do not have any mass current, i.e. it
is a state where the two spin species carry equal mass currents in the
opposite direction. We find that a mass current produces a stability
diagram identical to that of the spinless condensate in
Sec.~\ref{sec:chargecurrent}. From now on we will concentrate on the
case of spin twist only.

With a spin twist imposed on $\left|\Omega\right\rangle $, we expand
the Hamiltonian around its stationary state and investigate the
behavior of the fluctuation Hamiltonian. The full derivation is
carried out in Appendix~\ref{sec:App-spinful}. Here we shall present
the results, concentrating on the case where the spin current is put
along the diagonal: $\mathbf{p}=\tilde{p}\sum_{i}\hat{x}_{i}$.  For
comparison with the Bogoliubov theory, we plot in the right column of
Fig.~\ref{fig:pkphase} the wavenumber of the unstable wavevectors
(parallel to the spin current) as a function of the spin twist for
different interaction strengths. We find that, contrary to the
Bogoliubov theory, the region of unstable wavevectors decreases with
increasing interaction within the Gutzwiller formalism. For example,
the state at $U/U_c=0.7$ with $\tilde{p}>\pi/2$ is stable in the
Gutzwiller formalism while it shows instability within the Bogoliubov
theory. The Bogoliubov theory, which is accurate in the weakly
interacting regime, thus overestimates the dynamical instability in the
intermediate regime.The main qualitative difference, however, is in
the stability of the Neel state ($\tilde{p}=\pi$). While the
Bogoliubov theory predicts only a LI and no DI for this state, the
Gutzwiller ansatz shows that the Neel state can be dynamically
unstable in the intermediate interaction regime.

\begin{figure*}
\begin{centering}
\includegraphics[width=17cm]{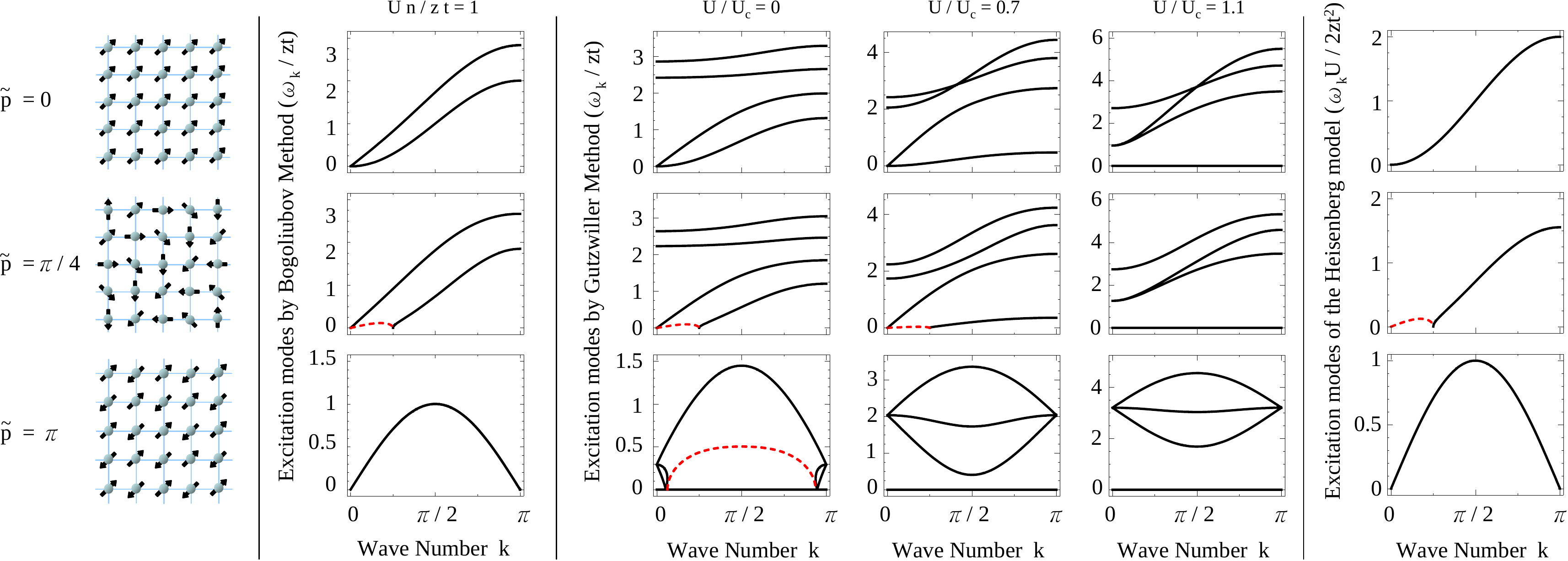}
\par\end{centering}

\caption{Dispersions of excitation modes for different spin twists along diagonal $\mathbf{p}=\tilde{p}(\hat{x}+\hat{y})$
where $\tilde{p}=0,\frac{\pi}{4}$ and $\pi$ obtained with Bogoliubov theory, Gutzwiller
approach, and within Heisenberg model. Solid lines are modes with
real energies while the red dashed lines are the imaginary part of
complex energies, which appears in conjugate pairs but only the positive
imaginary part is plotted. Note the energies of collective modes from Heisenberg model are in units of $2zt^2/U$ while those from Bogoliubov theory and Gutzwiller ansatz are in units of $zt$. \label{fig:Dispersions}}
\end{figure*}

In the middle column of Fig.~\ref{fig:Dispersions} we plot the
dispersion of the low energy collective modes (with
$\mathbf{k}\parallel\mathbf{p}$) of the ferromagnetic state, the Neel
state and a spin-spiral state with a period of $4$ lattice spacings,
for different interaction strengths. The ferromagnetic state has two
gapless modes in the weakly interacting limit: a linearly dispersing
charge mode and a quadratically dispersing spin mode. As interaction
strength is increased towards the critical interaction for the
superfluid-insulator transition, $U_c$, the charge mode dispersion is
almost unaffected, while the spin mode dispersion flattens out. Beyond
the critical coupling, in the Mott phase, the charge mode is gapped
out while the zero energy spin mode becomes dispersionless. This is an
artifact of our variational approach and we will discuss in the next
section how this degeneracy can be lifted by considering the
super-exchange mechanism of spin fluctuations. As soon as a spin
current is imposed (say for the spin-spiral state), the spin mode
develops a DI near $k=0$ in the superfluid phase. Beyond the critical
coupling, the DI vanishes in the Gutzwiller approach and we recover
the non-dispersing spin mode. In the weakly interacting limit, the
Neel state develops a dynamical instability for collective modes around
$k=\pi/2$. This dynamical instability however vanishes before the Mott
transition point is reached.

Comparing the results from the Gutzwiller ansatz to those from the Bogoliubov
theory, we find that for a given $\tilde{p}$, the discrepancy between
the two theories increases with $U/t$, while for a given $U/t$, the
discrepancy increases with increasing $\tilde{p}$. This is understood
from the fact that the effective Mott boundary in presence of spin
currents is given by $U=U_c\gamma_p$, and so, increasing the pitch of
the spin-twist pushes the system closer to the Mott phase, where the
validity of the Bogoliubov theory is suspect.

\begin{figure}
\begin{centering}
\includegraphics[width=7cm]{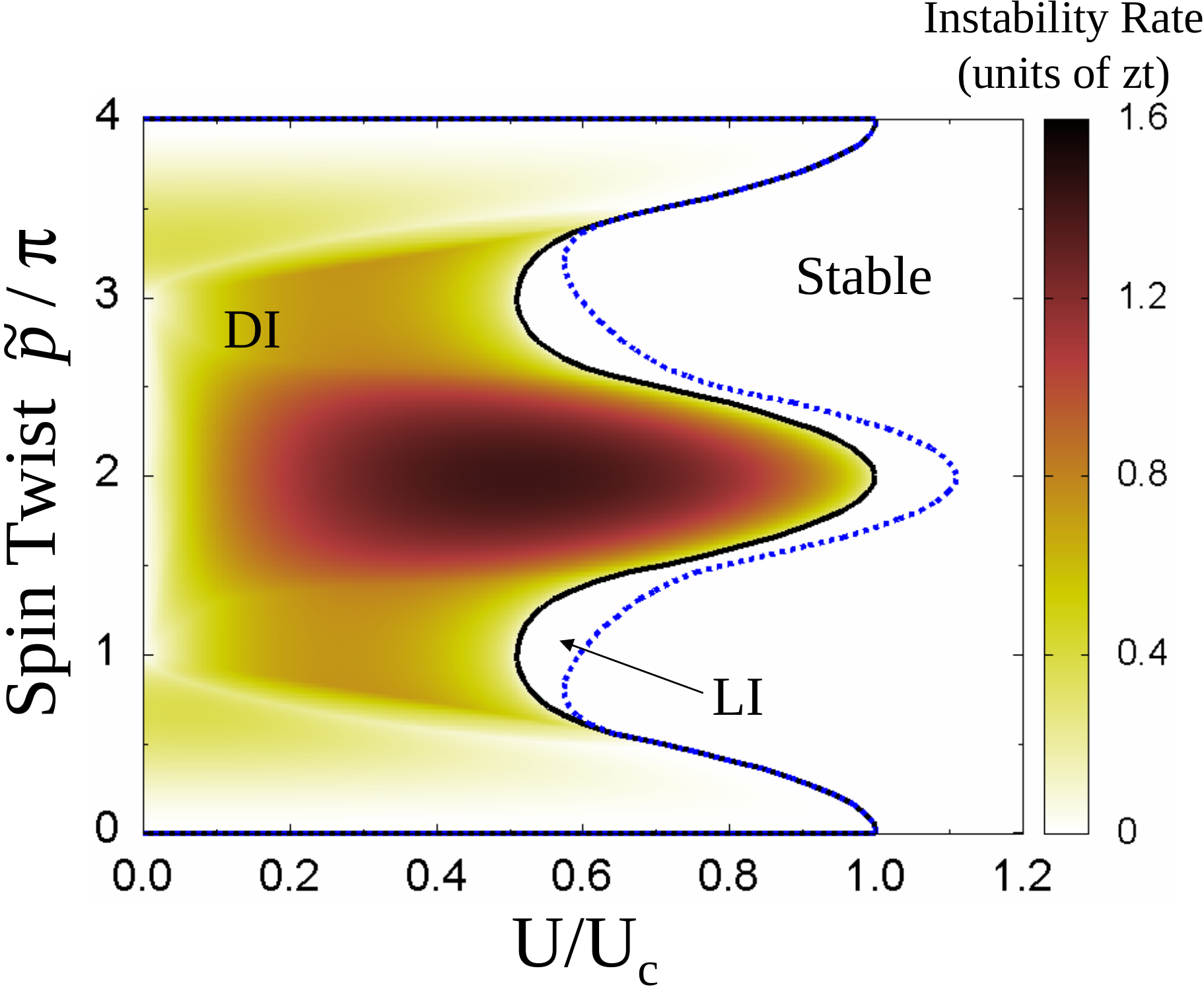}
\par\end{centering}

\caption{Phase Diagram of spin current on a lattice. This is computed with
$\gamma_{\mathbf{p}}=\cos \tilde{p}$ which represents a spin current put
along the diagonal of a square lattice. The shading indicates the
growth rate of the most unstable mode in the dynamically unstable
regime. \label{fig:spinfulPhase} }
\end{figure}

Fig.~\ref{fig:spinfulPhase} is the stability phase diagram of the two
component bosons in the interaction-spin-twist plane. We see that any finite
spin-twist leads to DI in the weakly interacting regime,
whereas, for $U>U_c/2$, states with $\tilde{p}$ around $\pi$
(including the Neel state) becomes stable. The color scale in the plot
represents the growth rate of the most unstable fluctuation mode in
the dynamically unstable region. The spin $1/2$ nature of the
particles is evident in the asymmetry of the growth rate between
$\tilde{p}=0$ and $\tilde{p}=2\pi$. Due to Berry's phase effects the
system is only symmetric under $4\pi$ (and not $2\pi$) twist of the
spin phase.

As in the case of mass current in Sec.~\ref{sec:chargecurrent}, the
validity of our Gutzwiller approach, and hence the results of
Fig.~\ref{fig:spinfulPhase}, are correct only if the fluctuation
occupation ($\sum_{n>0}d_{ni}^{\dagger}d_{ni}$ in
Appendix~\ref{sec:App-spinful}) is small. In the 2D case, we find that
it is indeed small ($<0.1$) for the majority of the stable regime, but
quickly goes up near the DI boundary, which is expected as a precursor
of instability. The most severe case happens at the DI boundary for
the Neel state, having fluctuation occupation $\sim 0.4$.

\subsection{Heisenberg Model \label{sub:Heisenberg-Model}}

It was previously shown that the two-component Bose-Hubbard model
(\ref{eq:2BH}) reduces to a ferromagnetic spin model in the deep
Mott phase \cite{Duan2003Controlling,Altman2003Phase}. Here we shall
show that within the Gutzwiller ansatz, the ferromagnetic ordering
is provided by the fluctuations.

First note that since a Mott phase has $\theta=0$ in our Gutzwiller
ansatz, all spin twists $\mathbf{p}$ give the same variational
ground state energy. The correction to the ground state energy due to
fluctuations is $\Delta E=\frac{1}{2}\sum_{k}\left(-\mathrm{Tr}M+\sum_{\alpha}\omega_{\alpha k}\right)$
where $M$ is the matrix derived in Appendix~\ref{sec:App-spinful}
and $\omega_{\alpha k}$ are the eigenenergies. We expand $\omega_{\alpha k}$
to the first order in $1/U$ and find that the correction to the ground
state energy for different spin twists obeys $E\left(\mathbf{p}\right)-E\left(0\right)=-\frac{Nt^{2}}{U}\left(\gamma_{\mathbf{p}}-1\right)$.
This is exactly the energy difference due to different magnons
in a ferromagnetic Heisenberg model.

However, one should be careful in interpreting the stability in the
deep Mott regime. As noted previously \cite{Altman2003Phase}, we
find a non-dispersing zero mode in the Mott phase, which emerges because
there is no energy cost to create a spin-flip locally. We note that
this is not physical and will be lifted at the next order in
perturbation theory.
Another way to understand this is that the product form we chose for
the variational state {[}Eq.~(\ref{eq:spinfulGutz}){]} is unable
to capture the spin ordering in the Mott phase, because charge fluctuations
are completely absent. To account for this, one can rotate the
state with a suitable unitary transformation $\left|G'\right\rangle =e^{iS}\left|G\right\rangle $,
which amounts to a canonical transformation $H'=e^{-iS}He^{iS}$ on
the Hamiltonian \cite{MacDonald1988$tU$,Chernyshev2004Higher}. To
the lowest order, $H$ acquires the correction $-J\sum_{\left\langle ij\right\rangle }\mathbf{S}_{i}\cdot\mathbf{S}_{j}$
where $J=\frac{4t^{2}}{U}$. Using this Hamiltonian in our analysis
we find that the zero mode is indeed lifted, with energy
\begin{equation}
\omega_{\mathbf{k}}(\mathbf{p})=\frac{zJ}{2}\sqrt{\left(\gamma_{\mathbf{p}}-\gamma_{\mathbf{k}}\right)\left(\gamma_{\mathbf{p}}-\frac{1}{2}\left(\gamma_{\mathbf{k}+\mathbf{p}}+\gamma_{\mathbf{k}-\mathbf{p}}\right)\right)}
\end{equation}
which is plotted in the right column in Fig.~\ref{fig:Dispersions}.
Note that this dispersion is identical to the usual spin mode in the
Heisenberg model with spin twist $\mathbf{p}$. Similar to the Bogoliubov results,
this spin mode has a LI for non-zero pitch and a DI for any pitch except
for the ferromagnetic and Neel state. However a crucial difference is
that the growth rate of the unstable modes in this case has
order of magnitude $\frac{t^{2}}{U}$, which is much smaller than
that of the DI we find in Fig.~\ref{fig:spinfulPhase}. This would
imply that the deep Mott state is at least \emph{quasi-stable}, in
that the instability time scale could be much longer than the experimental
time scale.

\section{Instabilities of the Neel state\label{sec:Discussion}}

From the beginning of implementation of optical lattices, observing
antiferromagnetically ordered states has been a holy grail of cold
atom experiments. Although the original ideas involved looking for
antiferromagnetic states with fermions, recently two component bosons
have been proposed as an alternate medium to observe
antiferromagnetism. In this context, there is a special interest in
the observation of Neel state with a commensurate spin-ordering vector
$\mathbf{p}=(\pi,\pi)$, which is notoriously hard to realize as a
ground state in cold atom systems
\cite{Lee2007Sublattice,Ho2008Intrinsic,Sorensen2010Adiabatic}.  In
the deep Mott phase, this state is the highest energy state of the
ferromagnetic spin model, and is expected to be stable
\cite{Purcell1951Nuclear,Ramsey1956Thermodynamics,Sorensen2010Adiabatic}
over relatively large time-scales, which has led to the idea that the
physics of the Neel state may be accessed in systems which are
carefully prepared to be stuck in this metastable state. In the
opposite limit of the weakly interacting superfluid phase, analysis
using the Bogoliubov approach in Sec.~\ref{sec:spinfulGP} also
demonstrates that the Neel state is dynamically stable. This naturally
leads to the question: Is the Neel state stable throughout the phase
diagram (i.e. for all interaction strengths)?

\begin{figure}
\begin{centering}
\includegraphics[width=6cm]{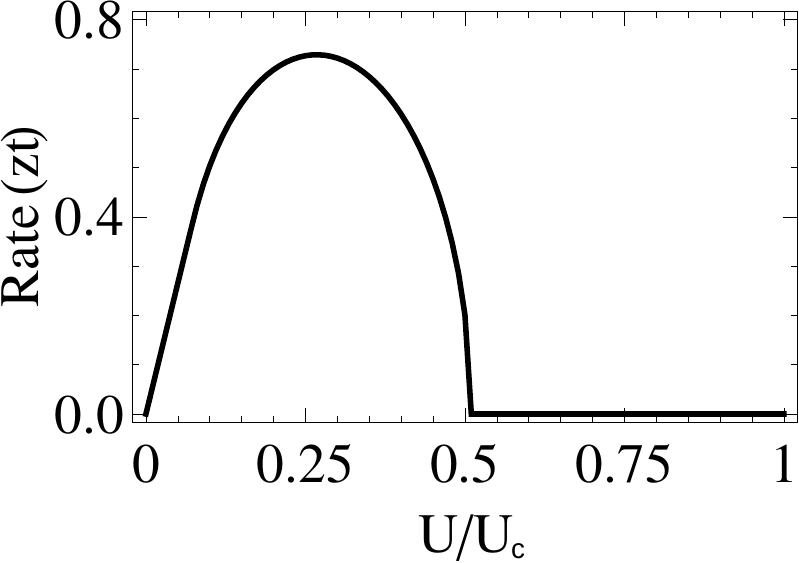}
\par\end{centering}
\caption{Growth rate of the most unstable fluctuation mode of Neel state as
a function of $U$, the on-site interaction, analyzed via the Gutzwiller approach. The result is expected to be invalid near $U=0$ because it is in superfluid phase.\label{fig:NeelInstability}}
\end{figure}

We use the Gutzwiller ansatz scheme to look at the stability of the
Neel state in the intermediate interaction regime. The Gutzwiller
approach shows that the Neel state is dynamically unstable for $0<U
\lesssim 0.51$. The $U=0$ state is technically stable, but is mostly
irrelevant for real experimental purposes as non-interacting bosons
are pathological even in equilibrium (e.g. divergent compressibility)
and need a finite interaction to form a stable superfluid.  In
Fig.~\ref{fig:NeelInstability}, we plot the growth rate of the most
unstable fluctuation mode (if any) of the Neel state, obtained via our
Gutzwiller approach as a function of the interaction strength. The
growth rate initially increases with the interaction strength in the
weakly interacting limit reaching a peak at around $U/U_c \sim
0.3$. It then decreases with increasing interaction and vanishes at
around $U/U_c \sim 0.5$. Thus Neel state physics can only be probed
with dynamically generated states for $U/U_c \gtrsim 0.5$. We note
that at $U/U_c=0.55$, the Gutzwiller ansatz predicts a condensate
depletion of about $0.21$, which shows that the approximation, which
involves a truncated Hilbert space, captures the essential physics in
this regime. The Gutzwiller results in the very weakly interacting
regime, $U\rightarrow 0$, are, however, suspect as the large number
fluctuations in this limit are incompatible with the truncation of the
Hilbert space used in the Gutzwiller scheme. In fact, to leading order
in the interaction strength, the Bogoliubov theory, which predicts a
stable Neel state, is much more trustworthy than the Gutzwiller
scheme. It would be interesting to see how the instability rates in
the Gutzwiller approximation change with increasing the size of the
Hilbert space, but this much more complicated problem is beyond the
scope of this paper.

\section{Conclusion\label{sec:Conclusion}}

In this work we have analyzed the stability of mass and spin current
carrying two-component Bose condensates in optical lattices. We have
approached the problem via Bogoliubov theory and the Gutzwiller
ansatz, to handle respectively the weakly interacting superfluid phase
and the regime near the Mott boundary. For small spin current and
small interaction the two approaches agree, but deviate when we
increase the spin current or the interaction strength. In the deep
Mott phase, we addressed the subtleties we encountered with our
variational approach, and showed that the ferromagnetic Heisenberg
model provides an adequate description in this limit.

For mass current carrying states, we find that the stability phase
diagram of the two component bosons qualitatively follow that of the
spinless bosons. The current carrying states are dynamically unstable
beyond a critical value of the current, and the critical current
monotonically decreases with increase in interaction strength,
vanishing at the critical coupling for the superfluid-insulator
transition. For the spin-current carrying states, we find, within
Bogoliubov theory, that the system is unstable to any finite spin
current in the weakly interacting limit, with the exception of the
Neel state (with a spin twist of $(\pi,\pi)$. The Gutzwiller approach
also predicts a similar scenario with the only difference being that
the Neel state is also dynamically unstable in this regime. The
Gutzwiller approach shows that the region of instability in the
spin-current-interaction plane decreases with increasing interaction,
with states around the Neel state (spin twist close to $(\pi,\pi)$)
being the dynamically stable states. The Heisenberg model in the
atomic limit also predicts dynamical instability for all current
carrying states except the Neel state. Finally we stressed the fact
that although the Neel state is stable in the two extreme limits of
strong and weak interactions, it can develop instabilities for
intermediate interaction strengths.

Although energetic instabilities would be very hard to observe
experimentally in cold atom systems, the dynamical instabilities of the
current carrying states should be fairly easily observable as dramatic
phenomena. The typical experiment would consist of creating a
spin-current carrying state by tuning a spatially varying artificial
Zeeman field. Such fields with commensurate wavevectors have already
been produced in the laboratories. By tuning the amplitude of these
fields to a very large value, so that the Zeeman energy is the largest
energy in the problem, the spin-current carrying states can be
generated as the ground states of the system of bosons. Once the field
is turned off, the system would exhibit violent disruption of the spin
pattern if it is in a dynamically unstable state, as long as the
inverse growth rate of the unstable modes are small compared to
experimental timescales. Since the maximum growth rate of unstable
modes is $\sim 1.5 zt$, this growth dynamics should be observable over
a wide range of experimental parameters. We thus hope that our
predicted instabilities would be easily seen by future experiments on
cold atoms.
 
\begin{acknowledgments}
We acknowledge useful discussions with J. V. Porto. This work was
supported by the NSF Joint Quantum Institute Physics Frontier Center.
\end{acknowledgments}
\appendix

\section{Analysis for Spinless Condensate \label{sec:App-spinless}}

Here we shall give the details of calculations in Sec.~\ref{sec:chargecurrent}.
The variational energy using the three-state Gutzwiller ansatz with phase twist $\mathbf{p}$ is:
\begin{eqnarray}
\epsilon & \equiv & \frac{1}{N}\left\langle \Omega\right|H\left|\Omega\right\rangle \nonumber \\
 & = & \left(\frac{U}{2}+\mu\cos\chi\right)\sin^{2}\frac{\theta}{2}-\frac{tz\bar{n}}{4}\gamma_{\mathbf{p}}\sin^{2}\theta\times\nonumber \\
 &  & \left(1+\bar{n}^{-1}\sin^{2}\frac{\chi}{2}+\sqrt{1+\bar{n}^{-1}}\sin\chi\cos2\eta\right)
\end{eqnarray}
where $\gamma_{\mathbf{p}}=z^{-1}\sum_{\mathbf{\delta}}e^{i\mathbf{p}.\mathbf{\delta}}$,
which reduces to $\gamma_{\mathbf{p}}=\frac{\cos p+(d-1)}{d}$ for a current
put along an axis of a $d$-dimension cubic lattice.

Since $\left|\Omega\right\rangle $ has filling ratio $\nu=\bar{n}-\sin^{2}\frac{\theta}{2}\cos\chi$,
to ensure commensurate filling one should find a $\mu$ such that
the minimum of $\epsilon$ occurs at $\chi=\frac{\pi}{2}$. For convenience
we consider only the limit $\bar{n}\gg1$. Then with $\mu=0$, the
minimum of $\epsilon$ is attained with $\chi=\frac{\pi}{2}$, $\eta=0$,
and
\begin{equation}
\cos\theta=\begin{cases}
u/\gamma_{\mathbf{p}}, & 0<u/\gamma_{\mathbf{p}}<1\\
1, & \mathrm{otherwise}
\end{cases}
\end{equation}
where $u\equiv\frac{U}{4zt\bar{n}}$. One can also check that this is
a stationary solution by varying $\epsilon$ locally to leading orders
to find $\delta\epsilon=\frac{1}{4}\left(U-4zt\bar{n}\gamma_{\mathbf{p}}\cos\theta\right)\sin\theta\delta\theta$.
The solution $\sin\theta=0$ corresponds to a Mott phase while the
other solution corresponds to a superfluid state.

Performing the fluctuation expansion as outlined in the text, one
finds $H=\frac{zt\bar{n}}{2}\sum_{\mathbf{k}}\Psi_{\mathbf{k}}^{\dagger}M_{\mathbf{k}}\Psi_{\mathbf{k}}$
where, in the superfluid and Mott phase, $M$ are respectively

\begin{widetext}

\begin{eqnarray}
M_{\mathrm{sf}} & = & \left(\begin{array}{cccc}
2\gamma_{\mathbf{p}}-\cos^{2}\theta\gamma_{+} & -\cos^{2}\theta\gamma_{+} & -\cos\theta\cos\frac{\theta}{2}\gamma_{-} & \cos\theta\cos\frac{\theta}{2}\gamma_{-}\\
-\cos^{2}\theta\gamma_{+} & 2\gamma_{\mathbf{p}}-\cos^{2}\theta\gamma_{+} & -\cos\theta\cos\frac{\theta}{2}\gamma_{-} & \cos\theta\cos\frac{\theta}{2}\gamma_{-}\\
-\cos\theta\cos\frac{\theta}{2}\gamma_{-} & -\cos\theta\cos\frac{\theta}{2}\gamma_{-} & \left(2\gamma_{\mathbf{p}}-\gamma_{+}\right)\cos^{2}\frac{\theta}{2} & \cos^{2}\frac{\theta}{2}\gamma_{+}\\
\cos\theta\cos\frac{\theta}{2}\gamma_{-} & \cos\theta\cos\frac{\theta}{2}\gamma_{-} & \cos^{2}\frac{\theta}{2}\gamma_{+} & \left(2\gamma_{\mathbf{p}}-\gamma_{+}\right)\cos^{2}\frac{\theta}{2}
\end{array}\right)\\
M_{\mathrm{mott}} & = & \left(\begin{array}{cccc}
2u-\gamma_{+} & -\gamma_{+} & -\gamma_{-} & \gamma_{-}\\
-\gamma_{+} & 2u-\gamma_{+} & -\gamma_{-} & \gamma_{-}\\
-\gamma_{-} & -\gamma_{-} & 2u-\gamma_{+} & \gamma_{+}\\
\gamma_{-} & \gamma_{-} & \gamma_{+} & 2u-\gamma_{+}
\end{array}\right)
\end{eqnarray}
\end{widetext}where $\gamma_{\pm}\equiv\frac{1}{2}\left(\gamma_{\mathbf{k}+\mathbf{p}}\pm\gamma_{\mathbf{k}-\mathbf{p}}\right)$.
Note that the problem reduces to the one considered before \cite{Altman2002Oscillating}
in the limit of $\mathbf{p}=0$.
The spectrum is found by diagonalizing $\sigma M$ where $\sigma\equiv\mathrm{diag}\left(1,-1,1,-1\right)$
which would give a spectrum of the form $\left\{ \pm\frac{1}{2}\omega_{\alpha\mathbf{k}}\right\} $.

\section{Analysis for Two-Component Condensate\label{sec:App-spinful}}

We take $\left|\Omega\right\rangle $ from Eq.~(\ref{eq:spinfulGutz})
with a spin twist $\mathbf{p}$ imposed to evaluate the variational energy $\epsilon=\frac{1}{N}\left\langle \Omega\right|H\left|\Omega\right\rangle $.
With some algebra it can be shown that one can first set
$\theta_{1,3,4}$ to
be $\pi/2$, after which
\begin{eqnarray}
\epsilon & = & \left(\frac{U}{2}+\mu\cos\theta_{2}\right)\sin^{2}\frac{\theta}{2}-\frac{tz\sin^{2}\theta}{8}\gamma_{\mathbf{p}/2}\nonumber \\
 &  & \times\left(3-\cos\theta_{2}+2\sqrt{2}\sin\theta_{2}\right).
\end{eqnarray}

To ensure a filling ratio $\nu=1$ one must set $\mu=-\frac{zt}{2}\gamma_{\mathbf{p}/2}\cos^{2}\frac{\theta}{2}$.
Then the variational energy is minimized by $\theta_{i>0}=\frac{\pi}{2}$
and
\begin{equation}
\cos\theta=\begin{cases}
u/\gamma_{\mathbf{p}/2} & ,0<u/\gamma_{\mathbf{p}/2}<1\\
1 & ,\mathrm{otherwise}
\end{cases}
\end{equation}
where $u\equiv\frac{U}{\left(3+2\sqrt{2}\right)zt}$. The six states
are written in terms of the E: $\left|\alpha\right\rangle \rightarrow t_{\alpha}^{\dagger}\left|\mathrm{vac}\right\rangle $
satisfying the constraint $\sum_{\alpha}t_{\alpha}^{\dagger}t_{\alpha}=1$,
where $\alpha$ could be the empty, 1 spin-up, 1 spin-down, 2 spin-up,
1 spin-up + 1 spin-down, or the 2 spin-down states. The boson creation/annihilation
operators are replaced by the pseudospin operators:
\begin{eqnarray}
b_{i\uparrow}^{\dagger} & \rightarrow & t_{\uparrow}^{\dagger}t_{0}+t_{\uparrow\downarrow}^{\dagger}t_{\downarrow}+\sqrt{2}t_{\uparrow\uparrow}^{\dagger}t_{\uparrow}\\
b_{i\downarrow}^{\dagger} & \rightarrow & t_{\downarrow}^{\dagger}t_{0}+t_{\uparrow\downarrow}^{\dagger}t_{\uparrow}+\sqrt{2}t_{\downarrow\downarrow}^{\dagger}t_{\downarrow}
\end{eqnarray}

The unitary transformation analogous to Eq.~(\ref{eq:spinlessUnitary})
is chosen as:\begin{widetext}
\begin{equation}
\left(\begin{array}{c}
d_{0i}^{\dagger}\\
d_{1i}^{\dagger}\\
d_{2i}^{\dagger}\\
d_{3i}^{\dagger}\\
d_{4i}^{\dagger}\\
d_{5i}^{\dagger}
\end{array}\right)=\left(\begin{array}{cccccc}
\frac{1}{\sqrt{2}}\sin\frac{\theta}{2} & \frac{e^{i\mathbf{p}\cdot\mathbf{x}_{i}/2}}{\sqrt{2}}\cos\frac{\theta}{2} & \frac{e^{-i\mathbf{p}\cdot\mathbf{x}_{i}/2}}{\sqrt{2}}\cos\frac{\theta}{2} & \frac{e^{i\mathbf{p}\cdot\mathbf{x}_{i}}}{2\sqrt{2}}\sin\frac{\theta}{2} & \frac{1}{2}\sin\frac{\theta}{2} & -\frac{e^{-i\mathbf{p}\cdot\mathbf{x}_{i}}}{2\sqrt{2}}\sin\frac{\theta}{2}\\
-\frac{1}{\sqrt{2}}\cos\frac{\theta}{2} & \frac{e^{i\mathbf{p}\cdot\mathbf{x}_{i}/2}}{\sqrt{2}}\sin\frac{\theta}{2} & \frac{e^{-i\mathbf{p}\cdot\mathbf{x}_{i}/2}}{\sqrt{2}}\sin\frac{\theta}{2} & -\frac{e^{i\mathbf{p}\cdot\mathbf{x}_{i}}}{2\sqrt{2}}\cos\frac{\theta}{2} & -\frac{1}{2}\cos\frac{\theta}{2} & -\frac{e^{-i\mathbf{p}\cdot\mathbf{x}_{i}}}{2\sqrt{2}}\cos\frac{\theta}{2}\\
\frac{1}{\sqrt{2}} & 0 & 0 & -\frac{e^{i\mathbf{p}\cdot\mathbf{x}_{i}}}{2\sqrt{2}} & -\frac{1}{2} & -\frac{e^{-i\mathbf{p}\cdot\mathbf{x}_{i}}}{2\sqrt{2}}\\
0 & \frac{e^{i\mathbf{p}\cdot\mathbf{x}_{i}/2}}{\sqrt{2}} & -\frac{e^{-i\mathbf{p}\cdot\mathbf{x}_{i}/2}}{\sqrt{2}} & 0 & 0 & 0\\
0 & 0 & 0 & \frac{e^{i\mathbf{p}\cdot\mathbf{x}_{i}}}{\sqrt{2}} & \frac{1}{\sqrt{2}} & \frac{e^{-i\mathbf{p}\cdot\mathbf{x}_{i}}}{\sqrt{2}}\\
0 & 0 & 0 & \frac{e^{i\mathbf{p}\cdot\mathbf{x}_{i}}}{\sqrt{2}} & 0 & -\frac{e^{-i\mathbf{p}\cdot\mathbf{x}_{i}}}{\sqrt{2}}
\end{array}\right)\left(\begin{array}{c}
t_{0i}^{\dagger}\\
t_{\uparrow i}^{\dagger}\\
t_{\downarrow i}^{\dagger}\\
t_{\uparrow\uparrow i}^{\dagger}\\
t_{\uparrow\downarrow i}^{\dagger}\\
t_{\downarrow\downarrow i}^{\dagger}
\end{array}\right)
\end{equation}
\end{widetext}and we set both $d_{0i}$ and $d_{0i}^{\dagger}$ to be
$\sqrt{1-\sum_{n>0}d_{ni}^{\dagger}d_{ni}}\approx1-\frac{1}{2}\sum_{n>0}d_{ni}^{\dagger}d_{ni}$
because $d_{0i}^{\dagger}$ is macroscopically occupied. The validity
of this expansion should be checked after the Bogoliubov transformation
to ensure consistency.

Written in terms of $d_{n>0}$ and to the lowest (quadratic) order,
$H=\frac{zt}{2}\sum_{\mathbf{k}}\left(\omega_{4\mathbf{k}}d_{4\mathbf{k}}^{\dagger}d_{4\mathbf{k}}+\Psi_{\mathbf{k}}^{\dagger}M\Psi_{\mathbf{k}}\right)$
where $\Psi_{\mathbf{k}}^{\dagger}\equiv\left(d_{1\mathbf{k}}^{\dagger},d_{1,-\mathbf{k}},d_{2\mathbf{k}}^{\dagger},d_{2,-\mathbf{k}},d_{3\mathbf{k}}^{\dagger},d_{3,-\mathbf{k}},d_{5\mathbf{k}}^{\dagger},d_{5,-\mathbf{k}}\right)$
and $M$ is an $8\times8$ matrix, whose non-zero entries are:
\begin{eqnarray*}
M_{1,1}=M_{2,2} & = & \frac{1}{2}\left(3+2\sqrt{2}\right)\left(u\cos\theta+\gamma_{\mathbf{p}/2}\sin^{2}\theta\right)\\
 &  & -\frac{1}{8}\gamma_{+}\left(9-2\sqrt{2}+\left(3+2\sqrt{2}\right)\cos2\theta\right)\\
M_{3,3}=M_{4,4} & = & \frac{1}{4}\left(3+2\sqrt{2}\right)\left(u+u\cos\theta+\gamma_{\mathbf{p}/2}\sin^{2}\theta\right)\\
 &  & -\frac{3}{2}\gamma_{+}\cos^{2}\frac{\theta}{2}\\
M_{5,5}=M_{6,6} & = & -\frac{1}{2}\left(3+2\sqrt{2}\right)(u-\gamma_{\mathbf{p}/2}(1+\cos\theta))\sin^{2}\frac{\theta}{2}\\
 &  & -\frac{1}{2}\gamma_{+}\sin^{2}\frac{\theta}{2}\\
M_{7,7}=M_{8,8} & = & \frac{1}{2}\cos^{2}\frac{\theta}{2}\left(\left(3+2\sqrt{2}\right)u-2\gamma_{+}\right)\\
 &  & +\frac{\gamma_{\mathbf{p}/2}}{2}\cos^{2}\frac{\theta}{2}\left(4+2\sqrt{2}-\left(3+2\sqrt{2}\right)\cos\theta\right)\\
M_{1,2}=M_{2,1} & = & -\frac{1}{8}\gamma_{+}\left(-3+6\sqrt{2}+\left(3+2\sqrt{2}\right)\cos2\theta\right)\\
M_{3,4}=M_{4,3} & = & \sqrt{2}\gamma_{+}\cos^{2}\frac{\theta}{2}\\
M_{1,3}=M_{3,1} & = & M_{2,4}=M_{4,2}=\frac{1}{2}\gamma_{\mathbf{p}/2}\cos\frac{\theta}{2}-\frac{1}{2}\gamma_{+}\cos^{3}\frac{\theta}{2}\\
M_{1,4}=M_{4,1} & = & M_{2,3}=M_{3,2}=\frac{1}{2}\gamma_{+}\cos\frac{\theta}{2}\sin^{2}\frac{\theta}{2}\\
M_{1,5}=M_{5,1} & = & -M_{2,6}=-M_{6,2}\\
 & = & \frac{1}{4}\gamma_{-}\left(-1+\sqrt{2}+\left(1+\sqrt{2}\right)\cos\theta\right)\sin\frac{\theta}{2}\\
M_{1,6}=M_{6,1} & = & -M_{2,5}=-M_{5,2}\\
 & = & \frac{1}{4}\gamma_{-}\left(-1+\sqrt{2}-\left(1+\sqrt{2}\right)\cos\theta\right)\sin\frac{\theta}{2}\\
M_{1,7}=M_{7,1} & = & -M_{2,8}=-M_{8,2}\\
 & = & \frac{1}{4}\gamma_{-}\cos\frac{\theta}{2}\left(2-\sqrt{2}+\left(2+\sqrt{2}\right)\cos\theta\right)\\
M_{1,8}=M_{8,1} & = & -M_{2,7}=-M_{7,2}\\
 & = & \frac{1}{4}\gamma_{-}\cos\frac{\theta}{2}\left(2-\sqrt{2}-\left(2+\sqrt{2}\right)\cos\theta\right)\\
M_{3,5}=M_{5,3} & = & -M_{4,6}=-M_{6,4}=\frac{\gamma_{-}}{2\sqrt{2}}\\
M_{3,6}=M_{6,3} & = & -M_{4,5}=-M_{5,4}=\frac{\gamma_{-}}{4}\sin\theta\\
M_{3,7}=M_{7,3} & = & -M_{4,8}=-M_{8,4}=\gamma_{-}\cos^{2}\frac{\theta}{2}\\
M_{3,8}=M_{8,3} & = & -M_{4,7}=-M_{7,4}=\frac{\gamma_{-}}{\sqrt{2}}\cos^{2}\frac{\theta}{2}\\
M_{5,7}=M_{7,5} & = & M_{6,8}=M_{8,6}\\
 & = & -\frac{1}{4}\left(2+\sqrt{2}\right)\gamma_{\mathbf{p}/2}\sin\theta-\frac{\gamma_{+}}{2\sqrt{2}}\sin\theta
\end{eqnarray*}
where $\gamma_{\pm}\equiv\frac{1}{2}\left(\gamma_{\mathbf{k}+\mathbf{p}/2}\pm\gamma_{\mathbf{k}-\mathbf{p}/2}\right)$.
Diagonalizing $\sigma M$ where $\sigma\equiv\mathrm{diag}\left(1,-1,1,-1,1,-1,1,-1\right)$
gives the spectrum.

\bibliography{Instability}

\end{document}